\documentclass{article}
\usepackage{spconf,amsmath,epsfig}
\usepackage{soul,color}
\usepackage{adjustbox}
\usepackage{amsmath}
\usepackage{amssymb}
\usepackage{amsthm}
\usepackage{algorithm}
\usepackage{algorithmic}
\usepackage{float}
\usepackage{graphicx}
\newcommand{\squeezeup}{\vspace{-2.5mm}}
\usepackage{hyperref}
\hypersetup{
    unicode=false,          % non-Latin characters in Acrobat’s bookmarks
    pdftoolbar=true,        % show Acrobat’s toolbar?
    pdfmenubar=true,        % show Acrobat’s menu?
    pdffitwindow=false,     % window fit to page when opened
    pdfstartview={FitH},    % fits the width of the page to the window
    pdftitle={},    % title
    pdfauthor={},     % author
    pdfsubject={},   % subject of the document
    pdfcreator={},   % creator of the document
    pdfproducer={}, % producer of the document
    pdfkeywords={Forward}{inverse}{modelling}{Gaussian process}{kernels}, % list of keywords
    pdfnewwindow=true,      % links in new windo
    colorlinks=true,       % false: boxed links; true: colored links
    linkcolor={blue},          % color of internal links (change box color with linkbordercolor)
    citecolor=blue,        % color of links to bibliography
    filecolor=blue,      % color of file links
    urlcolor=blue           % color of external links
}
\def\x{{\mathbf x}}

\title{Retrieval of Coloured Dissolved Organic Matter with Machine Learning Methods}

\name{Ana B. Ruescas$^1$, Martin Hieronymi$^2$, Sampsa Koponen$^3$, Kari Kallio$^3$  and Gustau Camps-Valls$^1$\thanks{The research was funded by the European Research Council (ERC) under the ERC-CoG-2014 SEDAL project (grant agreement 647423), and the Spanish Ministry of Economy and Competitiveness (MINECO) through the project TIN2015-64210-R. Especial thanks to Carsten Brockmann and the Case2eXtreme project team (funded by ESA).}}{}
\address{$^1$Image Processing Laboratory (IPL), Universitat de Val\`encia, Spain\\
$^2$ Institute of Coastal Rsearch, Helmholtz-Zentrum Geesthacht, Germany\\
$^3$Finnish Environment Institute SYKE, Finland}

\begin{document}
\onecolumn
%\ninept
%
\maketitle
\begin{abstract}
The coloured dissolved organic matter (CDOM) concentration is the standard measure of humic substance in natural waters. CDOM measurements by remote sensing is calculated using the absorption coefficient (\textit{a}) at a certain wavelength (e.g. $\approx 440 nm$). This paper presents a comparison of four machine learning methods for the retrieval of CDOM from remote sensing signals: regularized linear regression (RLR), random forest (RF), kernel ridge regression (KRR) and Gaussian process regression (GPR). Results are compared with the established polynomial regression algorithms. RLR is revealed as the simplest and most efficient method, followed closely by its nonlinear counterpart KRR.

\end{abstract}
\section{Introduction}
\label{sec:intro}
In the temperate and cold regions of the boreal zone humic waters are abundant, especially in lakes. %In Europe, boreal lakes mainly can be found in Finland, Sweden and Estonia. 
These waters typically have fairly low total suspended matter (TSM) and chlorophyll\_a (Chl-a) concentrations, even though  some cases of ``black lakes" with high Chl-a and TSM values have been reported too ~\cite{rs8060497}. For instance, in Finland the humic matter concentration of lakes correlates with the share of peat land in the drainage area ~\cite{doi:10.1139/f93-168}. Humic lakes can also originate from peat dredging, e.g. in the Netherlands. 
%The high CDOM concentration in humic waters increases attenuation of light in the blue and green region of the spectrum, and consequently decreases reflectance in the short wavelengths. 
CDOM absorption has an exponential shape and consequently decreases the water leaving reflectance in short wavelengths.

Information on humic substances is utilized in the application of official directives, lake management and climate change studies. Within the Water Framework Directive (WFD), in Sweden and Finland, water colour values $> 30mg$ Pt $l^{-1}$ (corresponds approximately $a_{CDOM}$(400) $3.6m^-1$) are defined as humic. Waters with water colour $> 90 mg$ Pt $l^{-1}$ ($a_{CDOM}$(400) $11 m^{-1}$) are classified as humus rich lakes~\cite{Koponen15}. An accurate measurement of the $a_{CDOM}$ parameter from remote sensing is crucial in these types of water. However, it is known that CDOM is the most critical and uncertain ocean colour (OC) product.
%In the present paper we focus on the CDOM estimation, showing results of the application of several machine learning (ML) algorithms in typical boreal waters, with medium to high CDOM concentrations ($1-20 m^{-1}$) and medium to low chlorophyll concentrations ($< 30 mg/m^{3}$). The data used is extracted from a Hydrolight simulation dataset prepared by ~\cite{Hieromyni2016} in the framework of the Case2eXtreme (C2X) project \footnote{\url{http://seom.esa.int/page_project014.php}}.

\begin{figure}[!ht]
\centering\includegraphics[width=0.8\linewidth]{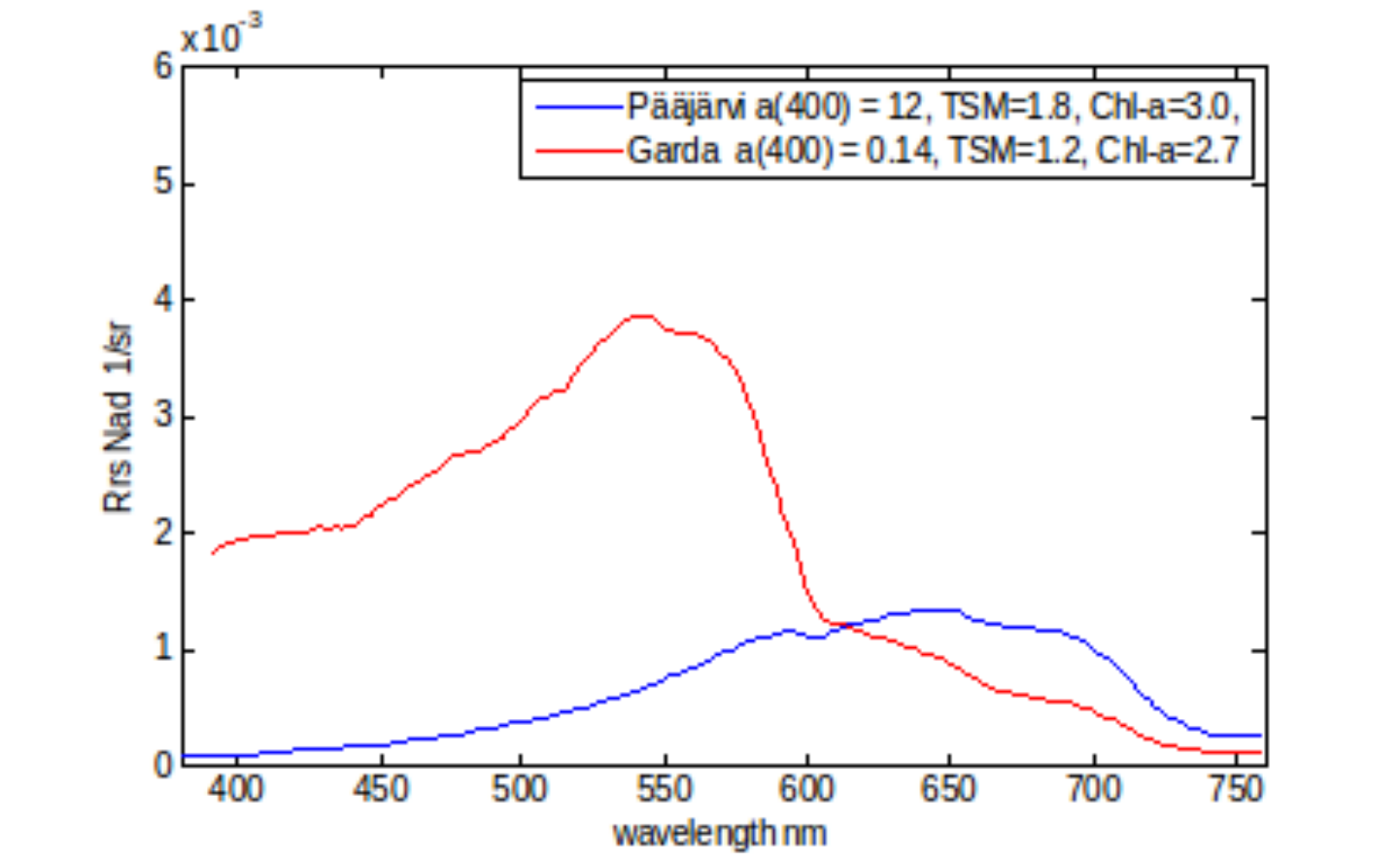}
\vspace*{-5mm}
\caption{\small Simulated $R_{rs}$ in a typical humic lake in Finland (P{\"a}{\"a}j{\"a}rvi) and in Lake Garda, Italy. $R_{rs}$ was simulated with the Hydrolight model \cite{Alikas14}} \label{Fig1}
\end{figure}

The impact of high CDOM on reflectance is demonstrated in Figure~\ref{Fig1}. In the present case, Lake Garda represents lake water with very low CDOM concentrations, while in Lake P{\"a}{\"a}j{\"a}rvi the CDOM concentration is high. The level of Chl-a is approximately the same in both lakes, and total suspended matter (TSM) is slightly lower in Lake Garda. $R_{rs}$ in the blue and green region of the spectrum is clearly higher in Lake Garda than in Lake P{\"a}{\"a}j{\"a}rvi. $R_{rs}$ at $<600 nm$ is significantly smaller in the CDOM rich lake. At $>650 nm$, where the influence of $a_{CDOM}$ is small, $R_{rs}$ in Lake Garda is slightly lower due to lower TSM concentration, and higher in Lake P{\"a}{\"a}j{\"a}rvi due to higher scattering of organic and inorganic particles.

%\squeezeup

Several band ratios have been proposed as predictive models for estimating CDOM from spectral data. These parametric approaches only take into account a few spectral bands and thus they disregard the information contained in other bands. Non-parametric regression algorithms can alternatively exploit the information contained in all spectral bands. In this paper, we assess the performance of several machine learning (ML) algorithms in CDOM estimation over typical boreal waters: multivariate regression, random forests, kernel ridge regression and Gaussian processes.%~\cite{CampsValls11mc}. 
We will use a Hydrolight simulation dataset presented in the framework of the Case2eXtreme (C2X) project\footnote{\url{http://seom.esa.int/page_project014.php}}~\cite{Hieromyni2016}. Basis for the tests are simulated hyperspectral $R_{rs}$ data that include extreme absorbing waters. We work with samples of medium to high CDOM concentrations at 440nm ($1-20 m^{-1}$) and medium to low Chl-a concentrations ($< 30 mg/m^{3}$). 

The remainder of the paper is organized as follows.
\S2 describes the methods used in this work.
\S3 gives an empirical evidence of performance of the proposed methods in comparison to standard bioptical models for the particular dataset used.
We conclude in \S4 with some remarks and an outline future work. 

\section{Methods and application}
\label{sec:ML}
\subsection{Established approaches with band ratio algorithms}
\label{sec:ratios}
Reports on band ratio algorithms are mostly based on airborne and field measurements with negligible or only small atmospheric influence. Important wavelength regions for the band ratio algorithms for CDOM retrieval are the ones between 400-600 nm taking 660-720 nm as reference. Following~\cite{Kallio183}~\cite{Brezonik2015199}, CDOM can be estimated by a ratio of reflectance at wavelength $> 600 nm$ to reflectance in the 400-550 nm range. This ratio is valid in a wide range of water constituent combinations. ~\cite{Alikas14}~\cite{Kallio183} used in situ measured data in Finland, and derived the algorithms that work well for $a_{CDOM}$ and TSM. The latter is estimated using a single band at 709 nm and not a ratio.  

\begin{figure}[!ht]
\centering\includegraphics[width=0.8\linewidth]{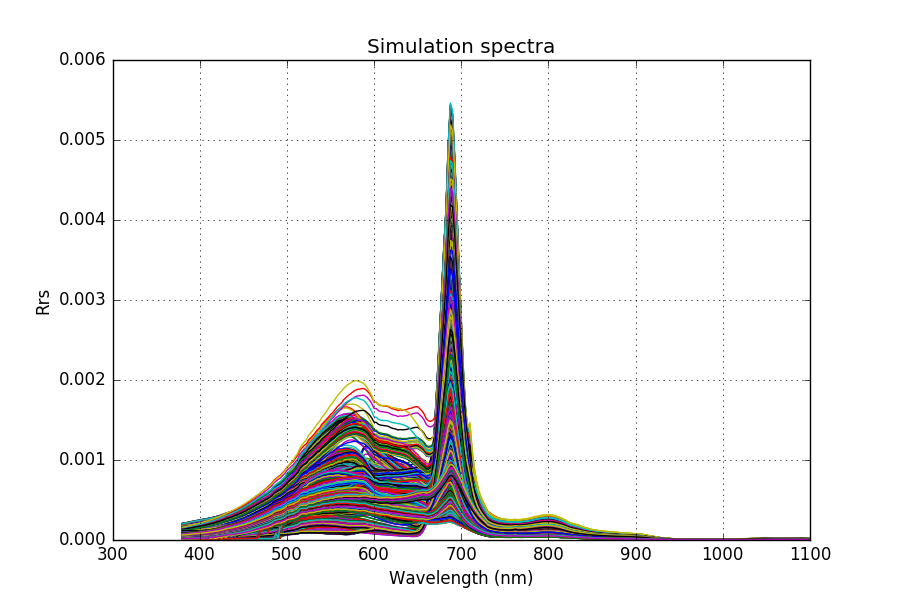}
\vspace*{-5mm}
\caption{\small The dataset includes the whole range of Chl-a ($0-200 mg/m^{3}$), with low TSM ($1-10  g/m^{3}$) and high CDOM ($1-20 m^{-1}$). When removing the Chl-a $>30 mg/m^{3}$, the fluorescence peak of the dataset at $\approx 700nm$ decreases considerably} \label{Fig2}
\end{figure}

The calibration for the CDOM algorithm with two band ratios is the main objective of Kallio's work~\cite{Kallio183}. He calculated Hydrolight simulations made with concentration data from monitoring stations in Finland (5553 in total) and used them for the calculation of the coefficients. Wavebands at 490, 665 and 709 nm (based on MERIS channels), were compared to the in situ CDOM measured at 440 nm. The optimal regression had a polynomial form with coefficients varying depending on the particular ratio used $x$:
\squeezeup
\begin{equation}
y = p_1*x^2 + p_2*x + p_3.
\end{equation}
The dataset used in this exercise comes from the simulated database of the C2X project and it is based on simulated remote sensing reflectance ($R_{rs}$), which is the ratio of water-leaving radiance to downwelling irradiance above the sea surface. Here, $R_{rs}$ refers to clear atmosphere with Sun at zenith and viewing angle exactly perpendicular. Highly absorbing waters are characterized by very low water-leaving radiance (black waters). The maximum of $R_{rs}$ is typically $< 0.005 sr^{-1}$ and located between 550 and 605 nm for Case-2 absorbing (C2A) cases. For extremely Case-2 absorbing waters (C2AX), the maximum shifts towards the red spectral range $> 600 nm$. The complete C2AX dataset used in the present work is illustrated in Figure \ref{Fig2}.

This complete dataset was filtered before using it here for Chl-a values $>30 mg/m^{3}$; and only nadir view angles have been left. Hihgh Chl-a values cuased the ratio to increase with low CDOM values and to decrease with high CDOM values. Those simulations were removed from the dataset since they are not frequent in nature. 

\subsection{Machine learning approaches}\label{sec:MLA}

Four machine learning algorithms for linear and non-linear regression are tested and compare to the polynomial regression explained above: (multivariate) linear regression (RLR), random forest (RF)~\cite{Breiman85}, kernel ridge regression (KRR)~\cite{Shawetaylor04}, Gaussian process regression (GPR)~\cite{Rasmussen06}. 

In multivariate (or multiple) linear regression (LR) the output $y$ (DCOM) is assumed to be a weighted sum of $B$ input variables, $\x:=[x_1,\ldots,x_B]^\top$, that is $\hat y=\x^\top{\bf w}$. Maximizing the likelihood is equivalent to minimizing the sum of squared errors, and hence one can estimate the weights ${\bf w}=[w_1,\ldots,w_B]^\top$ by least squares minimization. 
Very often one imposes some smoothness constraints to the model and also minimizes the weights energy, $\|{\bf w}\|^2$, thus leading to the regularized linear regression (RLR) method. 

An alternative method is that of random forests (RFs)~\cite{Breiman85}. A RF model is an ensemble learning method for regression that operates by constructing a multitude of decision trees at training time and outputting the mean prediction of the individual trees. The training algorithm for random forests applies the general technique of bootstrap aggregating, or bagging, to tree learners.

Kernel methods constitute a family of successful methods for regression~\cite{CampsValls09wiley}. We aim to incorporate two instantiations: (1) the KRR is considered as the (non-linear) version of the RLR~\cite{Shawetaylor04}; and (2) GPR is a probabilistic approximation to non-parametric kernel-based regression, where both a predictive mean and predictive variance can be derived~\cite{CampsValls16grsm}.
Kernel methods offer the same explicit form of the predictive model, which establishes a relation between the input (e.g., spectral data) $\x\in\mathbb{R}^{B}$ and the output variable (CDOM) is denoted as $y\in\mathbb{R}$. The prediction for a new radiance vector $\x_*$ can be obtained as:\\
\begin{equation}
\hat{y} = f(\x)= \sum\limits_{i=1}^{N} \alpha_i K_\theta (\x_i, \x_*) + \alpha_o,
\end{equation}
where $\{\x_i\}_{i=1}^{N}$ are the spectra used in the training phase, $\alpha_i$ is the weight assigned to each one of them, $\alpha_o$ is the bias in the regression function, and $K_\theta$ is a kernel or covariance function (parametrized by a set of hyperparameters $\boldsymbol{\theta}$) that evaluates the similarity between the test spectrum $\x_*$ and all $N$ training spectra. We used the automatic relevance determination (ARD) kernel function, 
%$K(\x_i,\x_j)=\nu\exp(-\|\x_i-\x_j\|^2/(2\sigma^2))+\sigma_n^2\delta_{ij}$, 
$$K(\x,\x')=\nu\exp\bigg(-\sum_{b=1}^B(x_b-x_b')^2/(2\sigma_b^2)\bigg)+\sigma_n^2\delta_{ij},$$ 
and learned the hyperparameters $\boldsymbol{\theta}=[\nu,\sigma_1,\ldots,\sigma_B,\sigma_n]$ by marginal likelihood maximization. An operational MATLAB toolbox is available at \url{http://isp.uv.es/soft_regression.html}. 

\subsection{An illustrative exercise}
In this exercise comparisons are made using the ratios as inputs, as well as the full spectral range found in the simulation dataset. This spectral range consisted in 6 bands corresponding to some of the MERIS \footnote{\url{https://earth.esa.int/web/guest/missions/esa-operational-eo-missions/envisat/instruments/meris}} sensor wavelengths: 442.5, 490, 510, 560, 665, 708.75 nm, from blue to near-infrared (NIR). This is the common set found in many of the ocean colour sensors used for water quality retrievals. Several trials using different inputs are tested: ratio 1 and ratio 2 are considered separately in the first two tests, both ratios are used together in the third test and all wavelengths are the multi-input variables in the fourth test. Statistics used to check the validity of the methods are: mean error (ME), root mean squared error (RMSE), normalized mean squared error (nMSE), mean absolute error (MAE) and the coefficient of correlation (R).

%%%%%%%%%%%%%%%%%%%%%%%%%%%%%%%%%%%%%%%%%%%%%%%%%%%%%%%
\section{Experimental Results}
\label{sec:simpleR}
%In this section we first ... and then. some remarks blablab
\subsection{Numerical comparison} 
The Table \ref{Tab:stats} offers an overview of the metrics for all four methods tested with the different input variables. When using only one input ($x_1 = R_{rs}(665)/R_{rs}(490nm)$, $x_2 = R_{rs}(709)/R_{rs}(490nm)$) the ML methods tested do not improve the results strikingly. The polynomial regression metrics are very similar to the others and especially close to the KRR method with ratio 1 ($x_1$) or the RF with ratio 2 ($x_2$). Even if in this second case the reduction in the ME and RSME is evident, computation time will then be the determining factor, and the polynomial regression requires less time and it still quite efficient.

\begin{table}[!ht]
\caption{\small Results obtained obtained with empirical fitting and several machine learning methods. Several scores are shown: mean error (ME), root-mean-square error (RMSE), mean absolute error (MAE) and Pearson's correlation coefficient (R).}
\centering
\small
\setlength{\tabcolsep}{3pt}
\begin{tabular}{|l|c|c|c|c|c|c|}
\hline
\textbf{} & \textbf{ME} & \textbf{RMSE} & \textbf{nMSE} & \textbf{MAE} & \textbf{R} \\
\hline
%\textbf{Ratio1} & & & & & \\
{\bf Ratio 1: $x_1 = 665/490$} & & & & &\\
\hline
Polyfit & 0.758 &3.832 & 0.319 & 2.940&0.570\\
RLR &0.624&4073 &-0.042 &3.174 & 0.438\\
RF &0.687 & 3.753& -0.078& 2.747& 0.592\\
KRR  &0.714& 3.707& -0.083& 2.767&0.586 \\
GPR &0.692&3.755&-0.078 &2.903 &0.585 \\
\hline
\textbf{Ratio 2: $x_2 = 709/490$} & & & & & \\
\hline
%\textbf{(709/490)} & & &  \\
Polyfit & 0.732 & 3.710& 0.263 & 2.846&0.603\\
RLR &0.634 & 3.969& -0.054& 3.071& 0.484\\
RF &0.387 & 3.320& -0.131& 2.336& 0.687\\
KRR  &0.648 & 3.411&-0.120&2.378& 0.665\\
GPR &0.808& 3.604& -0.096& 2.589& 0.638\\
\hline
\textbf{Both ratios: $\x=[x_1,x_2]$} & & & & & \\
\hline
%Polyfit & 0.758 &3.832 & 0.319 & 2.94&0.67\\
RLR &0.703& 3.900& -0.061& 2.956& 0.520\\
RF &0.363 & 2.296& -0.291& 1.497& 0.867\\
KRR  &0.588&2.842 & -0.199& 1.843& 0.802\\
GPR &0.487& 2.676& -0.225& 1.664 & 0.814\\
\hline
%\textbf{All reflectances, $\x=[x_1,\ldots,x_B]$} & & & & \\
\textbf{All bands, $\x\in{\mathbb R}^B$} & & & & & \\
\hline
%Polyfit & 0.758 &3.832 & 0.319 & 2.94&0.67\\
RLR & 0.259& 1.843& -0.387& 1.189& 0.913\\
RF & 0.137& 0.735& -0.786& 0.444& 0.987\\
KRR  & 0.045 & 0.812& -0.743& 0.453& 0.984\\
GPR & 0.014 & 0.646 & -0.842& 0.358& 0.990\\
\hline
\end{tabular}
\label{Tab:stats}
\end{table}

However, when using more than one variable, non-linear ML methods get more relevance, as it can be seen in the ``Both Ratios" and ``All bands" sections on the Table \ref{Tab:stats}. The RF seems to be the best model when using the two ratios together, while the GPR work well when using all the six reflectance bands. In this last case, the ``winner" method seems not to be so clear, since a simple RLR model improves already the results, although errors with RLR are double than using any other of the other three ML methods. 

\subsection{Analysis of the models}
\begin{figure}[!h]
\centering\includegraphics[width=0.9\linewidth]{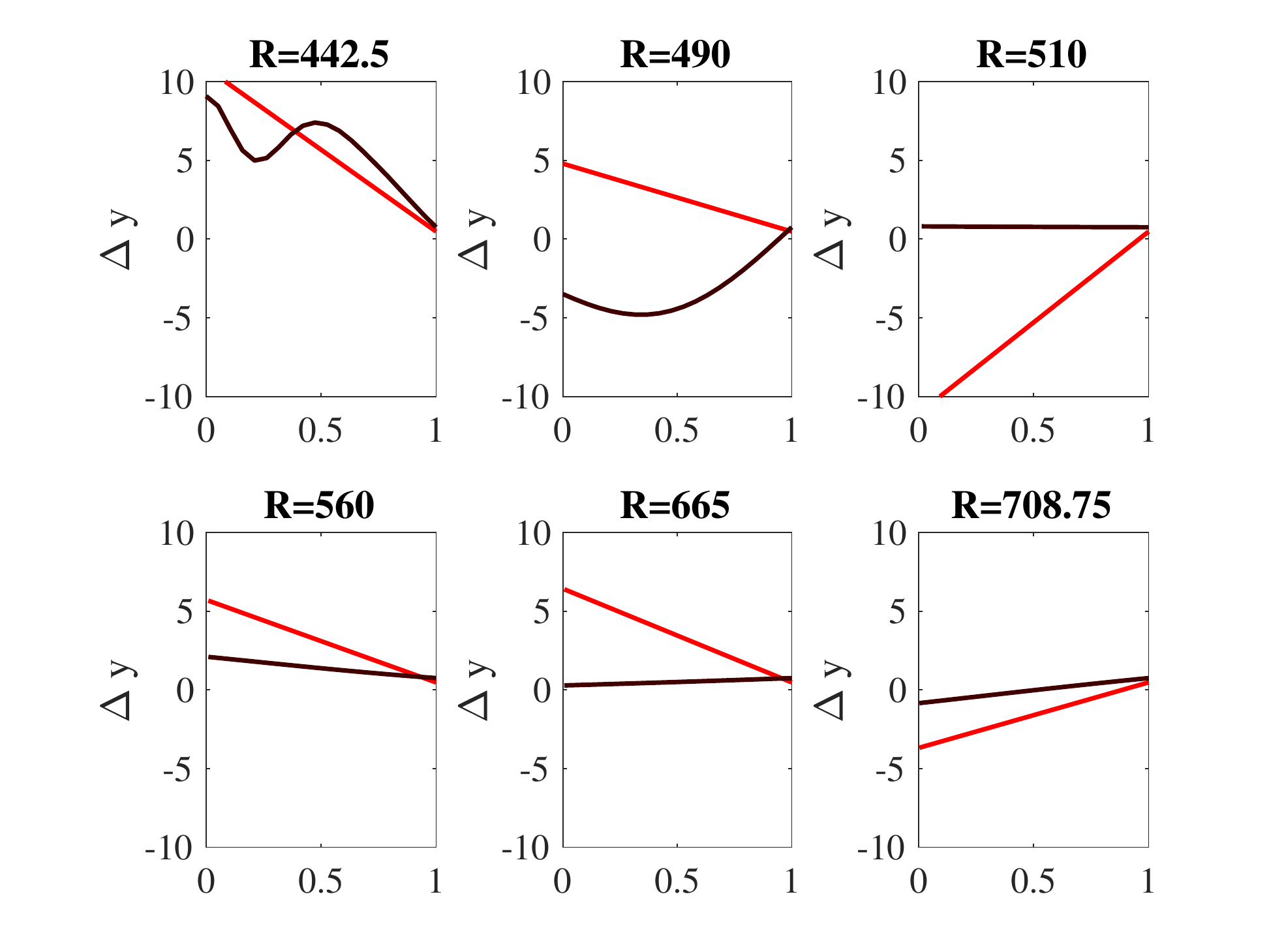}
\vspace*{-5mm}
\caption{\small Partial plots for the RLR (red) and the GPR (black) models.}\label{Fig3}
\end{figure}

Figure~\ref{Fig3} shows the partial dependence between the function (CDOM) and the target features (wavelengths) for the RLR model (red line) and the GPR model (black line). What can be inferred is that bands in the blue part of the spectrum (442.5, 490 nm) have a stronger influence in the CDOM derivation, while the green and red bands seem to be more neutral. The NIR band (709 nm) has a slight but determinant influence. 

\begin{figure}[!ht]
\centering\includegraphics[width=0.8\linewidth]{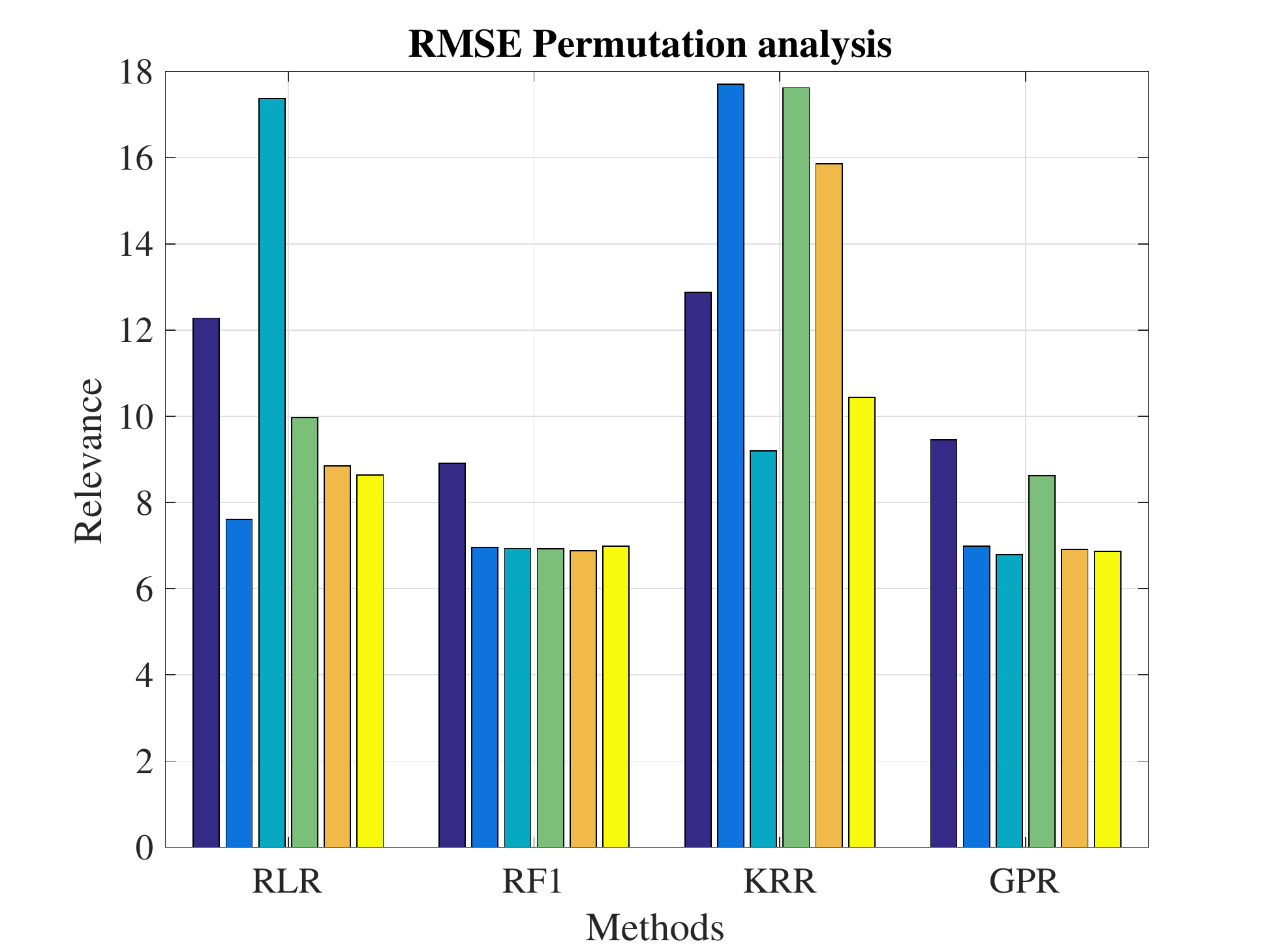}
\vspace*{-5mm}
\caption{\small RMSE permutation analysis of the models}\label{Fig4}
\end{figure}

A second analysis of the models considered a permutation analysis by which the impact of the input features on the prediction error is evaluated in the context or absence of the other predictors. Figure~\ref{Fig4} shows the influence of the wavelengths within the different models, which confirms the previous claim for all methods, with some variation in the KRR, where the weight of the green and red bands is higher. 

\subsection{Validation of the models}\label{sec:validation}
\begin{figure}[H]
\centerline{
\includegraphics[height=2cm]{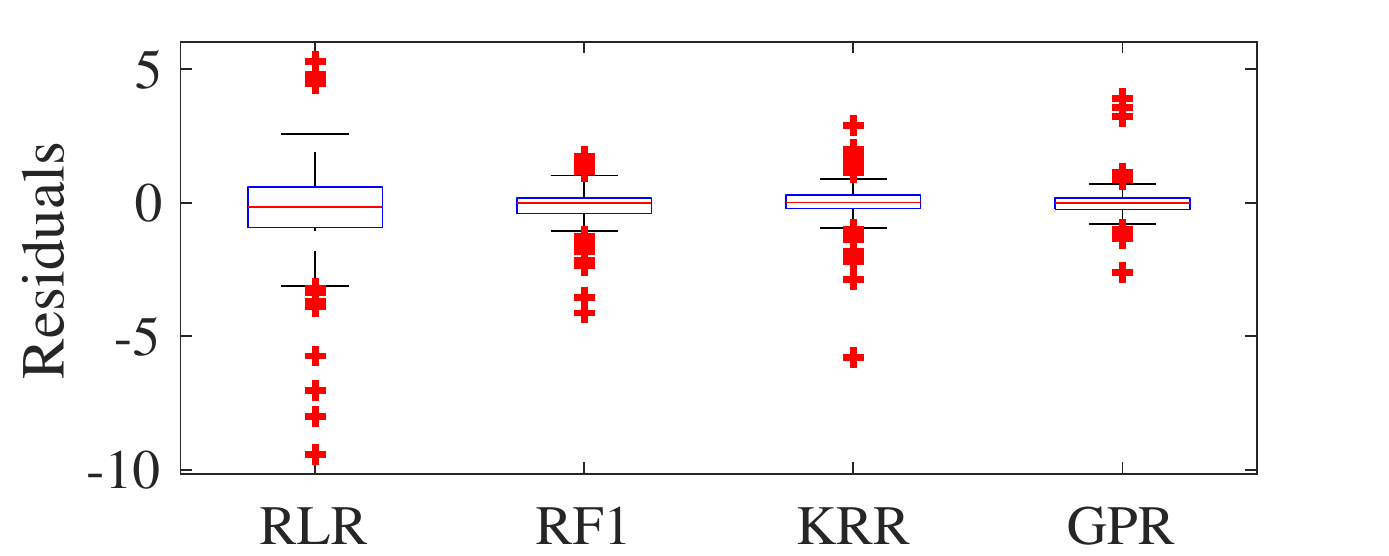}
\includegraphics[height=2cm]{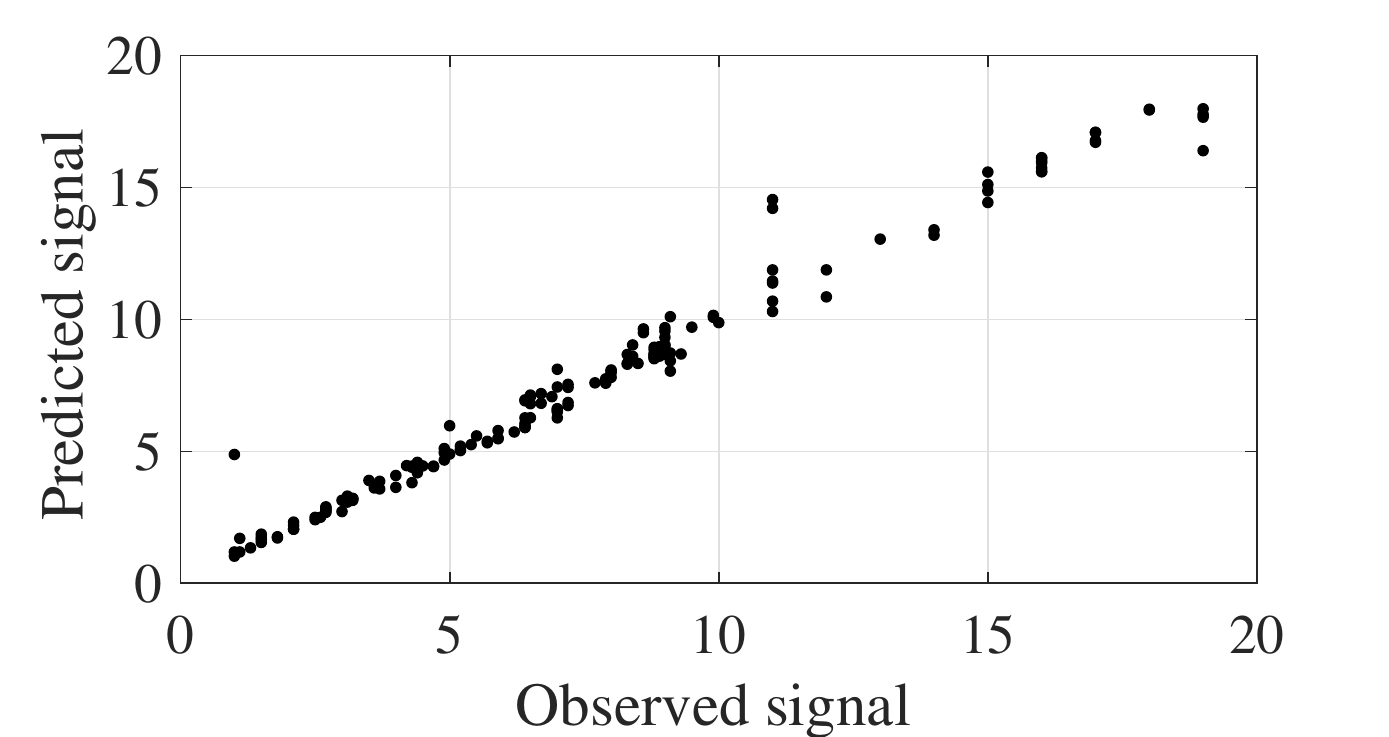}
}
\caption{\small Application of model on test data}\label{Fig5}
\end{figure}

One example of the application of the models to the test data and their validation is shown in Figure~\ref{Fig5}. Using all reflectance bands available, the plots show the summary of statistics of the residuals for the four methods (top) and the scatterplot of the linear regression between the observed and the predicted values for the method with the best RMSE result, in this case the GPR, which is also the best model in terms of ME and R (cf. Table \ref{Tab:stats}).

A similar exercise has been done with the retrieval of Chl-a (results not showed here). The dataset used was the C2A, with counted with a larger range of Chl-a values ($1-200 mg/m^{3}$) and still low TSM values ($0.2-10 g/m^{3}$). The summary statistics showed that the best performing method is the KRR but followed closely by the RLR, which improves considerably the simpler ratio method (709/665nm) e.g. from R=0.789 to R=0.902 or RMSE between 21.918 and 31.224 $mg/m^{3}$.

%To test the RLR method for CDOM and Chl-a with an external dataset would be the step forward. A validation with in situ CDOM and Chl-a measures is in the line of work as well.Similar results could also be tested for TSM retrieval in high scattering waters.

% \begin{table}[!ht]
% \caption{\small \textit{All reflectance} statistics comparison of the ML methods for the Chl-a variable compared with results of the ratio (709/665 nm)}
% \centering
% \small
% \setlength{\tabcolsep}{1.5 pt}
% \begin{tabular}{|l|c|c|c|c|c|c|}
% \hline
% \textbf{} & \textbf{ME} & \textbf{RMSE} & \textbf{nMSE} & \textbf{MAE} & \textbf{R} \\
% \hline
% %\textbf{Ratio1} & & & & & \\
% Ratio Chl & 0.439 &31.224 & -0.107 & 19.170&0.789\\
% RLR &0.733& 21.918 &-0.366 &14.035 & 0.902\\
% RF &0.827 & 18.906& -0.430& 8.778& 0.929\\
% KRR  &1.964& 18.634& -0.437& 8.885&0.932 \\
% GPR &6.573&33.007&-0.188 &14.539 &0.772 \\
% \hline
% \end{tabular}
% \label{Tab:stats_chl}
% \end{table}
\vspace*{-3mm}
%%%%%%%%%%%%%%%%%%%%%%%%%%%%%%%%%%%%%%%%%%%%%%%%%%%%%%%
\section{CONCLUSIONS}
\label{sec:con}
Four ML methods were tested using as training set simulated C2AX data, that is, extremely absorbing water with high CDOM concentration, Chl-a contents $<30 mg/m^{3}$ and low suspended matter. The traditional empirical algorithms are derived using band ratios and their correlation with measured or simulated $a_{CDOM}$. The polynomial algorithms derived using this empirical relationship are compared with more sophisticated ML methods using several variables as input: similar two-band ratios or the six MERIS like wavebands available. Results show that multivariate linear regression methods are already very efficient when using more than 2 bands or ratios, that is, more information coming from different parts of the spectrum. GPR methods give a very good result and are considered the best in terms of error and correlation, but the computation time needed increases considerably. 

% \section{REFERENCES}
% \label{sec:ref}

% List and number all bibliographical references at the end of the paper.  The references can be numbered in alphabetic order or in order of appearance in the document.  When referring to them in the text, type the corresponding reference number in square brackets as shown at the end of this sentence \cite{}.

\small
\bibliographystyle{IEEEbib}
\bibliography{CDOM.bib}

\end{document}